\documentclass[10pt, twocolumn, pre, aps, superscriptaddress, showpacs]{revtex4-1}
\usepackage{amsmath, graphicx, subfigure, tikz}
\begin{document}

\title{XY-Ashkin-Teller Phase Diagram in d=3}

\author{Alpar T\"urko\u{g}lu}
    \affiliation{Department of Physics, Bo\u{g}azi\c{c}i University, Bebek, Istanbul 34342, Turkey}
    \affiliation{Department of Electrical and Electronics Engineering, Bo\u{g}azi\c{c}i University, Bebek, Istanbul 34342, Turkey}
\author{A. Nihat Berker}
    \affiliation{Faculty of Engineering and Natural Sciences, Kadir Has University, Cibali, Istanbul 34083, Turkey}
    \affiliation{T\"UBITAK Research Institute for Basic Sciences, Gebze, Kocaeli 41470, Turkey}
    \affiliation{Department of Physics, Massachusetts Institute of Technology, Cambridge, Massachusetts 02139, USA}

\begin{abstract}
The phase diagram of the Ashkin-Tellerized XY model in spatial dimension $d=3$ is calculated by renormalization-group theory.  In this system, each site has two spins, each spin being an XY spin, that is having orientation continuously varying in $2\pi$ radians.  Nearest-neighbor sites are coupled by two-spin and four-spin interactions. The phase diagram has ordered phases that are ferromagnetic and antiferromagnetic in each of the spins, and phases that are ferromagnetic and antiferromagnetic in the multiplicative spin variable. The phase diagram exhibits two symmetrically situated reverse bifurcation points of the phase boundaries.  The renormalization-group flows are in terms of the doubly composite Fourier coefficients of the exponentiated energy of nearest-neighbor spins.
\end{abstract}
\maketitle

\section{Continuously Orientable Spins and Ashkin-Teller Complexity}

The conventional Ashkin-Teller model \cite{AT,Kadanoff0,Kecoglu} is a doubled-up Ising model, ushering a multiplicity of order parameters and ordered phases from this discrete-spin model.  Using continously orientable XY spins, instead of discrete Ising spins, brings even more interest.  The resulting model is defined by the Hamiltonian
\begin{multline}
- \beta {\cal H} = \sum_{\left<ij\right>} \, [J (\vec s_i \cdot \vec s_j +  \vec t_i \cdot \vec t_j) + M (\vec s_i \cdot \vec s_j)(\vec t_i \cdot \vec t_j) ]\\
= \sum_{\left<ij\right>} \, -\beta {\cal H}_{ij}(\vec s_i,\vec t_i;\vec s_j,\vec t_j)
\end{multline}
where $\beta=1/k_{B}T$ is the inverse temperature, at each site $i$ there are two XY unit spins $\vec s_i, \vec t_i$ that can point in $2\pi$ directions, and the sum is over all interacting quadruples of spins on nearest-neighbor pairs of sites.

\section{Method: Double Fourier Expansion of Two Continuous Angles}

The renormalization-group transformation, explained in Fig. 1, is done with length rescaling factor $b=3$ in order to conserve the ferromagnetic-antiferromagnetic symmetry of the method.  This method \cite{Migdal,Kadanoff} involves decimating three bonds in series into a single bond, followed by bond-moving by superimposing $b^{d-1}=9$ bonds.  This approach is an approximate solution on the $d=3$ cubic lattice and, simultaneously, an exact solution on the $d=3$ hierarchical lattice \cite{BerkerOstlund,Kaufman1,Kaufman2,BerkerMcKay}.  The simultaneous exact solution makes the approximate solution a physically realizable, therefore robust approximation, as also used in turbulence \cite{Kraichnan}, polymer \cite{Flory}, gel \cite{Kaufman}, electronic system \cite{Lloyd} calculations.  For recent works on hierarchical lattices, see Refs.\cite{Sponge,CubicSG,Clark,Kotorowicz,ZhangPP,Jiang,Derevyagin2,Chio,Teplyaev,Myshlyavtsev,Derevyagin,Shrock,Monthus,Sariyer}

\begin{figure}[ht!]
\centering
\includegraphics[scale=0.40]{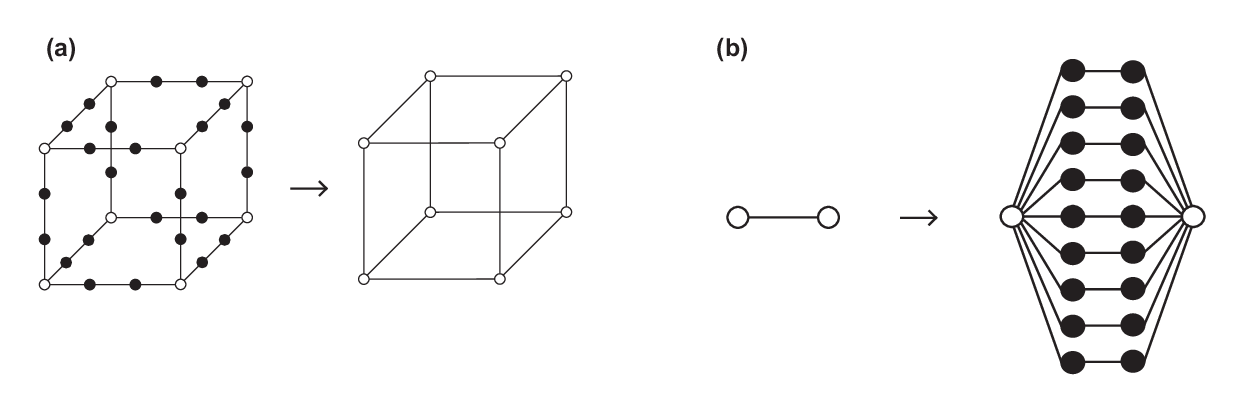}
\caption{(a) The Migdal-Kadanoff approximate renormalization-group transformation on the cubic lattice. Bonds are removed from the cubic lattice to make the renormalization-group transformation doable.  The removed bonds are compensated by adding their effect to the decimated remaining bonds.  (b) A hierarchical model is constructed by self-imbedding a graph into each of its bonds, \textit{ad infinitum}.\cite{BerkerOstlund}  The exact renormalization-group solution proceeds in the reverse direction, by summing over the internal spins shown with the dark circles.  Here is the most used, so called "diamond" hierarchical lattice \cite{BerkerOstlund,Kaufman1,Kaufman2,BerkerMcKay}.  The length-rescaling factor $b$ is the number of bonds in the shortest path between the external spins shown with the open circles, $b=3$ in this case. The volume rescaling factor $b^d$ is the number of bonds replaced by a single bond, $b^d=27$ in this case, so that $d=3$.}
\end{figure}

\begin{figure*}[ht!]
\centering
\includegraphics[scale=0.19]{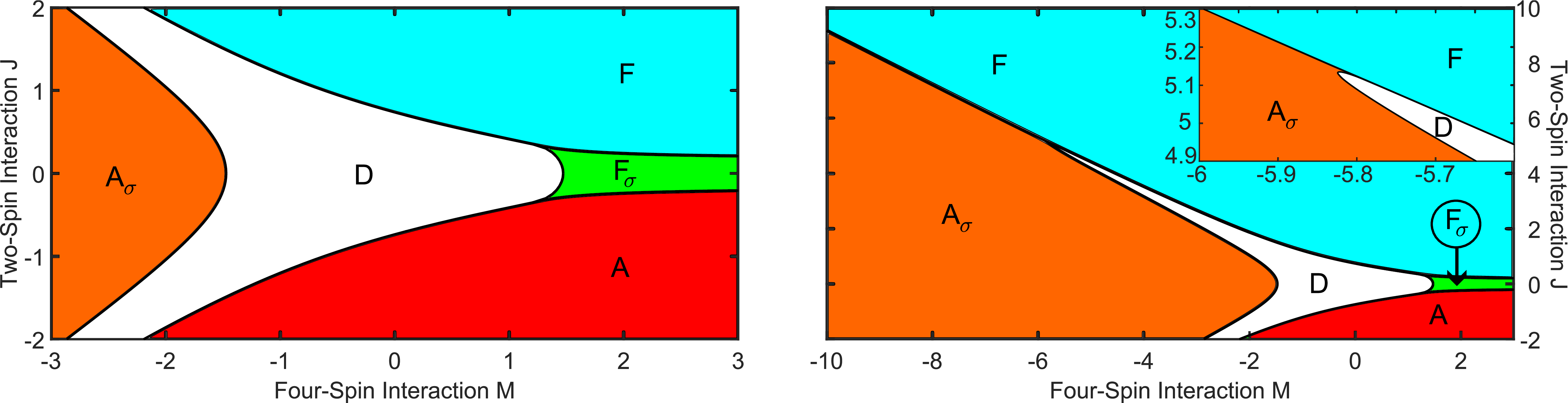}
\caption{Calculated phase diagram of the XY Ashkin-Teller model. The ferromagnetic $(F)$ and antiferromagnetic $(A)$ phases of the continuously orientable spin variables $\vec s_i$ and $\vec t_i$, and the ferromagnetic $(F_\sigma)$ and antiferromagnetic $(A_\sigma)$
phases of the composite spin variable $\vec s_i \vec t_i$, and the disordered phase $(D)$ are shown. The inset on the right shows the reverse bifurcation where a phase boundary splits into two phase boundaries and one of the latter reverses direction.  A similar phenomenon occurs symmetrically at negative $J$.}
\end{figure*}

As part of the first, decimation, step of the renormalization-group transformation, a decimated bond is obtained by integrating over the shared two spins of two bonds. With $u_{ij}(\theta_{ij},\varphi_{ij}) = e^{- \beta {\cal H}_{ij}(\vec s_i,\vec t_i;\vec s_j,\vec t_j)}$ being the exponentiated nearest-neighbor Hamiltonian between sites $(i,j)$, and $\theta_{ij}=\theta_i-\theta_j$ and $\varphi_{ij}=\varphi_i-\varphi_j$  being the angles between the planar unit vectors $(\vec s_i,\vec s_j)$ and $(\vec t_i,\vec t_j)$, decimation proceeds as
\begin{equation}
\tilde{u}_{13}(\theta_{13},\varphi_{13})=\int_0^{2\pi} u_{12}(\theta_{12},\varphi_{12}) u_{23}(\theta_{23},\varphi_{23}) \frac {d\theta_2}{2\pi} \frac {d\varphi_2}{2\pi}.
\end{equation}
Using the double Fourier transformation \cite{Jose,BerkerNel}
\begin{multline}
f(k,l) = \int_0^{2\pi}{u(\theta,\varphi)}e^{ik\theta+il\varphi} \frac {d\theta}{2\pi} \frac {d\varphi}{2\pi},\\
u(\theta,\varphi)=\sum_{k=-\infty}^{\infty} \sum_{l=-\infty}^{\infty}    e^{-ik\theta-il\varphi} f(k,l),
\end{multline}
the decimation of Eq.(2) becomes
\begin{equation}
\tilde{f}_{13}(k,l)=f_{12}(k,l)f_{23}(k,l).
\end{equation}

As part of the second, bond-moving, step of the renormalization-group transformation, a bond moving is effected as
\begin{multline}
u_{ij}'(\theta,\varphi)=\tilde{u}_{i_1j_1}(\theta,\varphi) \, \tilde{u}_{i_2j_2}(\theta,\varphi),\\
f_{ij}'(k,l)=\sum_{m=-\infty}^{\infty} \sum_{n=-\infty}^{\infty}  \tilde{f}_{i_1j_1}(k-m,l-n)  \tilde{f}_{i_2j_2}(m,n).
\end{multline}

We have followed the renormalization-group flows in terms of the $k=0$ to 20 and $l=0$ to 20 double Fourier components, also using $f_{ij}(k,l)=f_{ij}(-k,l)=f_{ij}(k,-l)=f_{ij}(-k,-l)$. We also made spot checks with higher number of double Fourier components (up to $k,l=0$ to 100). We set to unity the maximum value of the double Fourier components, by dividing with the same constant (the raw maximal value), which amounts to adding the same constant to all energies.

\begin{figure*}[ht!]
\centering
\includegraphics[scale=0.11]{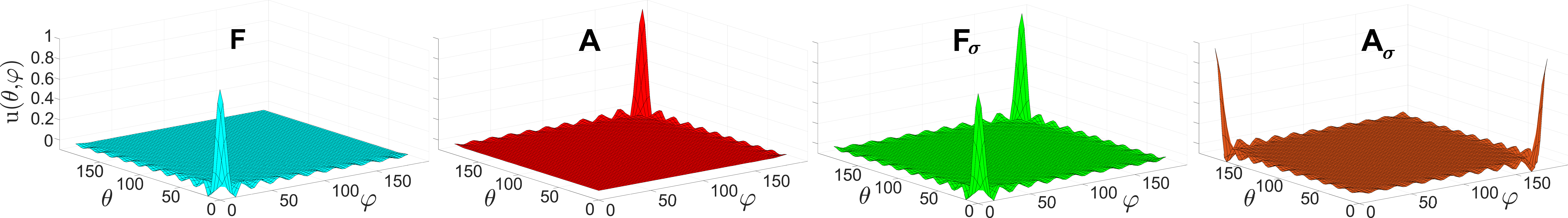}
\caption{Renormalization-group sinks of the phases of the XY Ashkin-Teller model. The exponentiated nearest-neighbor energy
$u_{ij} = e^{-\beta {\cal H}_{ij}(\vec s_i,\vec t_i;\vec s_j,\vec t_j)}$ is shown, as a function of the angle $\theta$ between the spins $\vec s_i,\vec s_j$ and the angle $\varphi$ between the spins $\vec t_i,\vec t_j$ . The energies are normalized to the maximum value of $u = 1$ by the inclusion of an additive constant to the energy.}
\end{figure*}

\section{Results: Phase Diagram, Entropic Ordered Phases, Reverse Bifurcation}
The phase diagram (Fig. 2) is obtained by following the renormalization-group flows of the double Fourier components, obtained as described above, to their stable fixed points, namely sinks.  The basin of attraction of each sink is a corresponding thermodynanic phase.\cite{BerkerWor}  In this XY-Ashkin-Teller model, there are five sinks and therefore five distinct thermodynamic phases.  The exponentiated nearest-neighbor interactions, $u_{ij}(\theta_{ij},\varphi_{ij}) = e^{- \beta {\cal H}_{ij}(\vec s_i,\vec t_i;\vec s_j,\vec t_j)}$, reconstructed [Eq.(3)] from the double Fourier coefficients, at four of these sinks are shown in Fig. 3.  A sink epitomizes the ordering of its corresponding thermodynamic phase that it attracts under renormalization group. Thus, as seen leftmost in Fig. 3, in the ferromagnetic phase $F$, the $\vec s_i$ spins are aligned $(\theta_{ij}=0)$ with each other and separately the $\vec t_i$ spins are aligned $(\varphi_{ij}=0)$ with each other.  In the antiferromagnetic phase $A$, the neighboring $\vec s_i$ spins are antialigned $(\theta_{ij}=\pi)$ with each other and separately the neighboring $\vec t_i$ spins are antialigned $(\varphi_{ij}=\pi)$ with each other.  In the entropic composite ferromagnetic phase $F_\sigma$, the neighboring spins $\vec s_i = \pm \vec s_j$ and simultaneously $\vec t_i = \pm \vec t_j$, the upper (or lower) signs being jointly valid, $\theta_{ij},\varphi_{ij}=0$ or $\pi$.  In the also entropic composite antiferromagnetic phase, in neighboring $(s_i,s_j)$ and $(t_i,t_j)$ are either respectively aligned and antialigned, or respectively antialigned and aligned. The latter two ordered phases have an entropy per bond $S/N = \ln 2.$  Furthermore, in all four ordered phases, the relative orientation of the $s_i$ and $t_i$ systems has a global degeneracy of $2\pi$.  The sink of the disordered phase (not shown in Fig. 3) has the (0,0) double Fourier component equal to unity, all other double Fourier components equal to zero.  Therefore, $u_{ij}(\theta_{ij},\varphi_{ij}) = e^{- \beta {\cal H}_{ij}(\vec s_i,\vec t_i;\vec s_j,\vec t_j)} = 1$ independent of angle. Calculated phase diagram exhibits (inset of Fig. 2) a reverse bifurcation where a phase boundary splits into two phase boundaries and one of the latter reverses direction.  This phenomenon occurs symmetrically at positive and negative $J$.

\section{Conclusion} 

We have solved, by renormalization-group theory, the Ashkin-Teller type doubled-up XY magnetic spin model in spatial dimension $d=3$. We find four different ordered phases, with ferromagnetic and antiferromagnetic orderings of the direct and composite spins. Two reverse bifurcations, where a phase boundary splits into two phase boundaries and one of the latter reverses directions, occurs symmetrically at positive and negative $J$.

\begin{acknowledgments} Support by the Academy of Sciences of Turkey (T\"UBA) is gratefully acknowledged.
\end{acknowledgments}

\end{document}